\documentclass[a4paper]{article}
\usepackage[dvips]{graphics,graphicx}
\usepackage{amsmath}
\begin{document}
\newcommand{\PRL}[1]{Phys.\ Rev.\ Lett. \textbf{#1}}
\newcommand{\PRD}[1]{Phys.\ Rev.\ D\,\textbf{#1}}
\newcommand{\PRC}[1]{Phys.\ Rev.\ C\,\textbf{#1}}
\newcommand{\PLB}[1]{Phys.\ Lett.\ B\, \textbf{#1}}
\newcommand{\PTP}[1]{Prog.\ Theor.\ Phys. \textbf{#1}}
\newcommand{\NPA}[1]{Nucl.\ Phys.\ \textbf{A#1}}
\newcommand{\EPJC}[1]{Eur.\ Phys.\ J.\ C\,\textbf{#1}}
\mark{{Analysis of one- and two-particle spectra...}{T.~Hirano, K.~Morita,
S.~Muroya, and C.~Nonaka}}
\title{Analysis of one- and two-particle spectra at RHIC based on a
hydrodynamical model}

\author{Tetsufumi Hirano$^1$, Kenji Morita$^2$\footnote{Presenter}, Shin
Muroya$^3$, and Chiho Nonaka$^4$ \\
$^1$Physics Department, University of Tokyo, Tokyo 113-0033, Japan\\
$^2$Department of Physics, Waseda University, Tokyo 169-8555, Japan\\
$^3$Tokuyama Women's College, Tokuyama, Yamaguchi 745-8511, Japan\\
$^4$IMC, Hiroshima University, Higashi-Hiroshima, Hiroshima 739-8521, Japan}
\date{}
\maketitle

\abstract{
 We calculate the one-particle hadronic spectra and correlation functions of
 pions based on a hydrodynamical model. Parameters in the model
 are so chosen that the one-particle spectra reproduce experimental results
 of $\sqrt{s}=130A$GeV Au+Au collisions at RHIC. Based on the numerical
 solution, we discuss the space-time evolution of the fluid.  Two-pion
 correlation functions are also discussed. Our numerical solution suggests
 the formation of the quark-gluon plasma with large volume and low net baryon
 density.}
\section{Introduction}
The new state of matter, the quark-gluon plasma (QGP), is expected to be
produced in relativistic heavy ion collisions. Though the recent experimental
data of $\sqrt{s}=130A$GeV Au+Au collisions at RHIC have already been
presented \cite{QM2001}, the collision process is too complicated to
understand clearly due to many-body interactions and multiparticle
productions. Therefore we need a simple phenomenological description as a
basis of the further investigation. In this talk, based on a
hydrodynamical-model calculation, we try to give a possible explanation for
the RHIC experimental results of the single-particle spectra and correlation
function \cite{paper}.

\section{Model and Particle Spectra}
Assuming that local thermal and chemical equilibrium are established, the
dynamical evolution of the system can be described by a relativistic
hydrodynamics. We numerically solve the hydrodynamical equations for the
perfect fluid and baryon number conservation with cylindrical symmetry
\cite{Ishii_PRD46}, where we take both longitudinal and transverse dynamical
expansion into account.  We use an equation of state (EoS) with a phase
transition of first order \cite{Nonaka_EPJC}. 

\begin{figure}[htbp]
\begin{center}
 \includegraphics[width=0.6\textwidth]{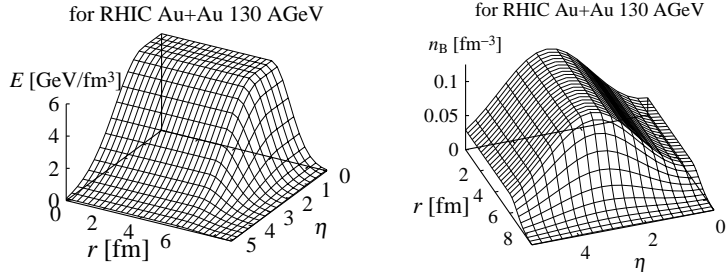}
 \caption{Left: Initial energy density
 distribution. Right: net baryon number distribution.}
 \label{fig:initial}
\end{center}
\end{figure}

We put the initial time as $\tau=\sqrt{t^2-z^2}=1.0$ fm. Initial longitudinal
velocity is assumed to be Bjorken's solution, $v_z=z/t$. Initial transverse
flow velocity is neglected. We parameterize the initial energy density
distribution and net baryon number distribution as in Fig.~\ref{fig:initial}.

For both RHIC and SPS, we adjust the parameters in the model so that
calculated single-particle spectra reproduce the experimental results.
The parameter sets for both experiments are summarized in 
Table.~\ref{tbl:param}. 

\begin{table}[ht]
 \begin{center}
  \caption{Parameters and output from the fluid.}
  \label{tbl:param}
  \begin{tabular}[t]{ccc}\hline\hline
   & Au+Au at 130 $A$GeV & Pb+Pb at 17.4 $A$GeV \\ \hline
   $E_{\text{max}}$ & 6.0 GeV/fm$^3$ & 5.74 GeV/fm$^3$ \\
   $n_{\text{max}}$& 0.125 fm$^{-3}$ & 0.7	fm$^{-3}$\\
   Total $n_{\text{B}}$ & 132 & 305 \\
   $\langle T_{\text{f}} \rangle $ & 125 MeV & 123 MeV \\
   $\langle \mu_{\text{B}} \rangle$ & 76.1 MeV & 280.2 MeV \\
   $\langle v_{\text{T}} \rangle_{|\eta|\leq 0.1}$ & 0.51 &
   0.467
   \\ \hline
  \end{tabular}
 \end{center}
\end{table}

Once the freeze-out hypersurface is fixed, one can calculate the
single-particle spectra via Cooper-Frye formula. 
We take account of not only hadrons directly emitted from the freeze-out
hypersurface but also the ones produced in the decay of resonances. Note that
the resonances are also emitted from the freeze-out hypersurface. 
We include decay processes $\rho\rightarrow 2\pi$, $\omega\rightarrow 3\pi$,
$\eta\rightarrow 3\pi$, $K^* \rightarrow \pi K$, and $\Delta\rightarrow
N\pi$. 

\begin{figure}[ht]
\begin{center}
 \includegraphics[width=0.65\textwidth]{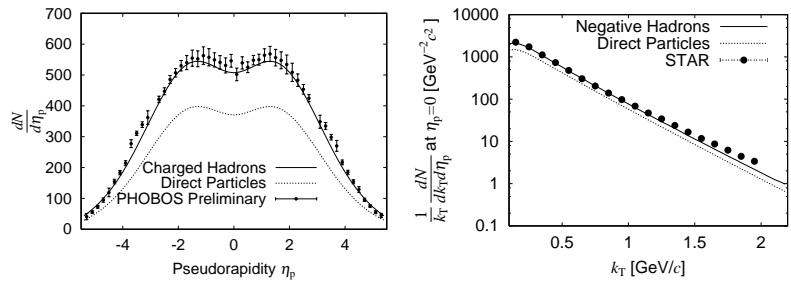}
 \caption{Single-particle spectra at RHIC. Left:
 Pseudorapidity distribution. Right: Transverse momentum distribution. In
 both figures, solid lines, dashed lines and symbols denote total number of
 particles (i.e., including resonance decay), direct components and
 experimental data, respectively.}
 \label{fig:spectra}
\end{center}
\end{figure}

Figure \ref{fig:spectra} shows pseudorapidity distribution of charged
hadrons and transverse momentum distribution of negatively charged hadrons. 
Preliminary experimental data in the pseudorapidity distribution are
taken from the PHOBOS Collaboration \cite{PHOBOS_QM2001}. Experimental data
in the $k_{\text{T}}$ distribution are taken from the STAR
Collaboration \cite{STAR_PT}. Our results well reproduce the experimental
data. 

The parameter set in Table.\ \ref{tbl:param} tells us that the energy density
is not extremely high but sufficient for the QGP production. The value 6.0
GeV/fm$^3$ is only 5\% higher than the one of SPS, 5.74 GeV/fm$^3$ at the
same initial time, $\tau=1.0$ fm. Our value is smaller than different
calculations \cite{Hirano_PRC}. This is due to the different
thermalization time and the different initial energy density profile. Major
discrepancy between the RHIC and the
SPS can be seen in the net baryon density and longitudinal extension of the
fluid. The net baryon number at the RHIC is much smaller than the one at the
SPS. Longitudinal extension along $\eta$-axis at the RHIC is twice as large
as at the SPS. Hence, much higher colliding energy at the RHIC is converted
into the large volume of the hot matter rather than high energy density.
We can also see that the transverse flow 
$v_{\text{T}}=\tanh Y_{\text{T}}$
becomes stronger at RHIC than at SPS.

It is well known as Hanbury-Brown Twiss effect that one can get the
information on the source of particles through two-particle correlation
function. Based on the numerical solution of the hydrodynamical equation, we
calculate the two-pion correlation functions. Here we consider only the
pions directly emitted from the hypersurface, for simplicity. The HBT radii
are extracted from the correlation function via a Gaussian fitting function
\begin{equation}
 C_{\text{2fit}}(K, q) = 1+\lambda
  \exp  \left[-R_{\text{out}}^2(K) 
		 q_{\text{out}}^2
		 -R_{\text{side}}^2(K) 
		 q_{\text{side}}^2 
		 -R_{\text{long}}^2(K) 
		 q_{\text{long}}^2 
		 \right].
\end{equation}

\begin{figure}[ht]
\begin{center}
 \includegraphics[width=0.8\textwidth]{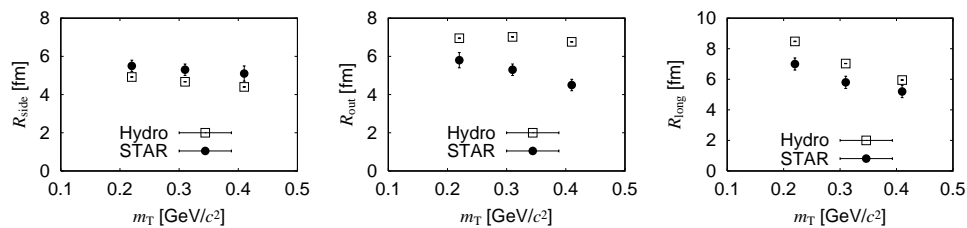}
 \caption{HBT radii at RHIC. From left to right, 
 $R_{\text{side}},  R_{\text{out}}$ and $R_{\text{long}}$,
 respectively. Experimental data are take from the
 STAR Collaboration \protect\cite{STAR_HBT}.}
 \label{fig:c2rhic}
\end{center}
\end{figure}

\begin{figure}[ht]
\begin{center}
 \includegraphics[scale=0.6]{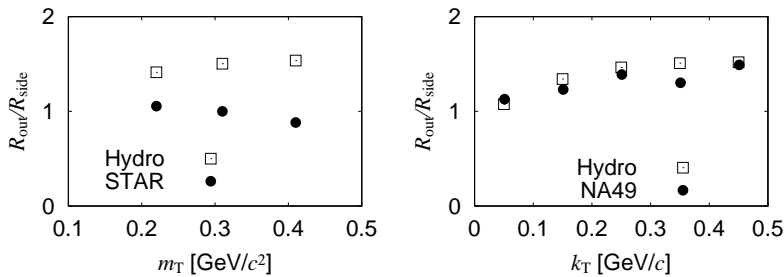}
 \caption{$R_{\text{out}}/R_{\text{side}}$. Left: RHIC. Right: SPS.}
 \label{fig:ratio}
\end{center}
\end{figure}

Transverse mass dependence of the HBT radii is displayed in
Fig. \ref{fig:c2rhic}. The sideward HBT radii, 
$R_{\text{side}}$, 
show almost consistent result with the experiment. 
The outward and longitudinal HBT radii are larger than
 the experimental data. The outward HBT radii
show different behavior with respect to 
$M_{\text{T}}$. 
The experimental result strongly depends on $M_{\text{T}}$ 
but our result does not show the strong dependence on
$M_{\text{T}}$. 
We also show the 
$R_{\text{out}}/R_{\text{side}}$ 
which should be a
good indicator of the long-lived fluid \cite{Rischke_NPA608} in
Fig.~\ref{fig:ratio}. Our calculation agrees with SPS data, in which the
ratio monotonically increases with $K_{\text{T}}$. As shown in
Table.~\ref{tbl:param}, Our numerical solution of the hydrodynamical
equations does not show major difference in lifetime between the RHIC and the
SPS because the initial energy density at RHIC is only slightly higher than
that at SPS. In fact, the time duration at RHIC is about 6 fm. Hence, the
ratio shows similar behavior both qualitatively and quantitatively in our
result. However, the experimental result at RHIC clearly decreases with
$M_{\text{T}}$ and the value is smaller than our
result. For the definite conclusion of this puzzle, we need more and
more theoretical investigation.

\section{Summary}
In summary, we analyze the one-particle distribution and HBT radii based on a
hydrodynamical model. Our model successfully reproduce the one-particle
distribution at RHIC experiment. The numerical solution shows the QGP
production with low net baryon number density and large volume. The maximum
energy density is not much higher than the SPS case. HBT radii are also
investigated. Our result in Fig.~\ref{fig:c2rhic} shows roughly agreement
with the experiment. Figure \ref{fig:ratio} may indicate different particle
emission mechanism at RHIC.

The author would like to thank Drs. I. Ohba, H. Nakazato, H. Nakamura and
P. Kolb for their helpful discussion and comments. This work is partially
 supported by Waseda University Grant for Special Research Projects
 No.2001A-888 and Waseda University Media Network Center.


\begin{thebibliography}{99}
 \bibitem{QM2001} Quark Matter 2001, to be published in the proceedings of the
		 Fifteenth International Conference on Ultra-Relativistic
		 Nucleus-Nucleus Collisions. (Stony Brook, USA, January 14-21, 2001)

 \bibitem{paper}T. Hirano, K. Morita, S. Muroya, and C. Nonaka,
		 nucl-th/0110009. See also references therein.

 \bibitem{Ishii_PRD46}T. Ishii and S. Muroya, \PRD{46}, 5156 (1992).

 \bibitem{Nonaka_EPJC}C. Nonaka, E. Honda, and S. Muroya, \EPJC{17}, 663
		 (2000).

 \bibitem{PHOBOS_QM2001}A. Wuosmaa \textit{et al.} (PHOBOS Collaboration), 
		 in \cite{QM2001}.

 \bibitem{STAR_PT}C. Adler \textit{et al.} (STAR Collaboration), \PRL{87},
		 112303 (2001).

 \bibitem{Hirano_PRC}See e.g., T. Hirano, \PRC{65}, 011901(R) (2002).

 \bibitem{STAR_HBT}C. Adler \textit{et al.} (STAR Collaboration), \PRL{87},
		 082301 (2001).

 \bibitem{Rischke_NPA608}D. H. Rischke and M. Gyulassy, \NPA{608}, 479
		 (1996).
\end{thebibliography}
\end{document}